\documentstyle[osa,10pt]{revtex}
\begin{document}

\title{Quantum jumps in backaction evasion measurements 
of a light field component}

\author{Holger F. Hofmann and Takayoshi Kobayashi \\
Department of Physics, Faculty of Science, University of Tokyo,\\
7-3-1 Hongo, Bunkyo-ku, Tokyo113-0033, Japan\\
Akira Furusawa\\Nikon Corporation, R\&D Headquarters,\\
Nishi-Ohi, Shinagawa-ku, Tokyo 140-8601, Japan}

\maketitle

\begin{abstract}
Backaction evasion measurements of a quadrature component of the light field 
vacuum necessarily induce quantum jumps in the photon number. The correlation 
between measurement results and quantum jump events reveals fundamental 
nonclassical aspects of quantization.\\
(OCIS codes:(270.0270) Quantum Optics; (270.5290) Photon statistics)
\end{abstract}
\vspace{0.5cm}
In optical backaction evasion quantum nondemolition measurements, 
the nonlinear interaction
between a meter light field and the signal field is applied to obtain 
information about the physical properties of the signal without absorbing
it \cite{Fri92,Per94}. Nevertheless the uncertainty principle requires a 
nonvanishing backaction on those light field variables which do not 
commute with the observed signal property. By studying this unavoidable
backaction, fundamental insights can be gained on the quantum mechanical 
relationship between non-commuting properties. In particular, the field 
nature of photon number quantization may be investigated by correlating 
phase dependent field properties with the photon number of a light field 
state \cite{Hof00a,Hof00b}.  

\begin{figure}[h]
\setlength{\unitlength}{1.2pt}
\begin{picture}(400,240)

\put(35,60){\line(1,1){50}}
\put(45,60){\line(1,1){45}}
\put(90,110){\line(0,-1){10}}
\put(90,110){\line(-1,0){10}}

\put(15,46){\makebox(40,12){\large Meter}}
\put(15,34){\makebox(40,12){\large input}}

\put(35,180){\line(1,-1){50}}
\put(45,180){\line(1,-1){45}}
\put(90,130){\line(0,1){10}}
\put(90,130){\line(-1,0){10}}

\put(15,196){\makebox(40,12){\large Signal}}
\put(15,184){\makebox(40,12){\large input}}

\put(70,120){\line(1,0){60}}
\put(20,120){\makebox(40,12){\large Beam}} 
\put(20,108){\makebox(40,12){\large splitter}}

\put(105,130){\line(1,1){45}}
\put(115,130){\line(1,1){40}}
\put(155,175){\line(0,-1){10}}
\put(155,175){\line(-1,0){10}}

\put(105,110){\line(1,-1){45}}
\put(115,110){\line(1,-1){40}}
\put(155,65){\line(0,1){10}}
\put(155,65){\line(-1,0){10}}

\put(160,25){\makebox(80,15){\Large $x \to a x$}}
\put(160,10){\makebox(80,15){\Large $y \to y/a$}}
\put(160,45){\framebox(80,40){\Large OPA}}
\put(160,155){\framebox(80,40){\Large OPA}}
\put(160,215){\makebox(80,15){\Large $x \to x/a$}}
\put(160,200){\makebox(80,15){\Large $y \to a y$}}

\put(245,70){\line(1,1){40}}
\put(245,60){\line(1,1){45}}
\put(290,110){\line(0,-1){10}}
\put(290,110){\line(-1,0){10}}

\put(245,170){\line(1,-1){40}}
\put(245,180){\line(1,-1){45}}
\put(290,130){\line(0,1){10}}
\put(290,130){\line(-1,0){10}}

\put(270,120){\line(1,0){60}}
\put(340,120){\makebox(40,12){\large Beam}} 
\put(340,108){\makebox(40,12){\large splitter}}

\put(305,130){\line(1,1){45}}
\put(315,130){\line(1,1){40}}
\put(355,175){\line(0,-1){10}}
\put(355,175){\line(-1,0){10}}

\put(305,110){\line(1,-1){45}}
\put(315,110){\line(1,-1){40}}
\put(355,65){\line(0,1){10}}
\put(355,65){\line(-1,0){10}}

\put(340,196){\makebox(40,12){\large Meter}}
\put(340,184){\makebox(40,12){\large output}}

\put(340,46){\makebox(40,12){\large Signal}}
\put(340,34){\makebox(40,12){\large output}}

\end{picture}
\setlength{\unitlength}{1.2pt}
\caption{\label{setup} Schematic setup for the quantum nondemolition 
measurement of the quadrature component $\hat{x}_M$ of the signal field.
Note that the beam splitter reflectivity $R$ must be adjusted to 
correspond to the amplification factor $a$ of the OPAs. Its value should 
be equal to $R=a^2/(a^2+1)$.}
\vspace{0.5cm}
\end{figure}
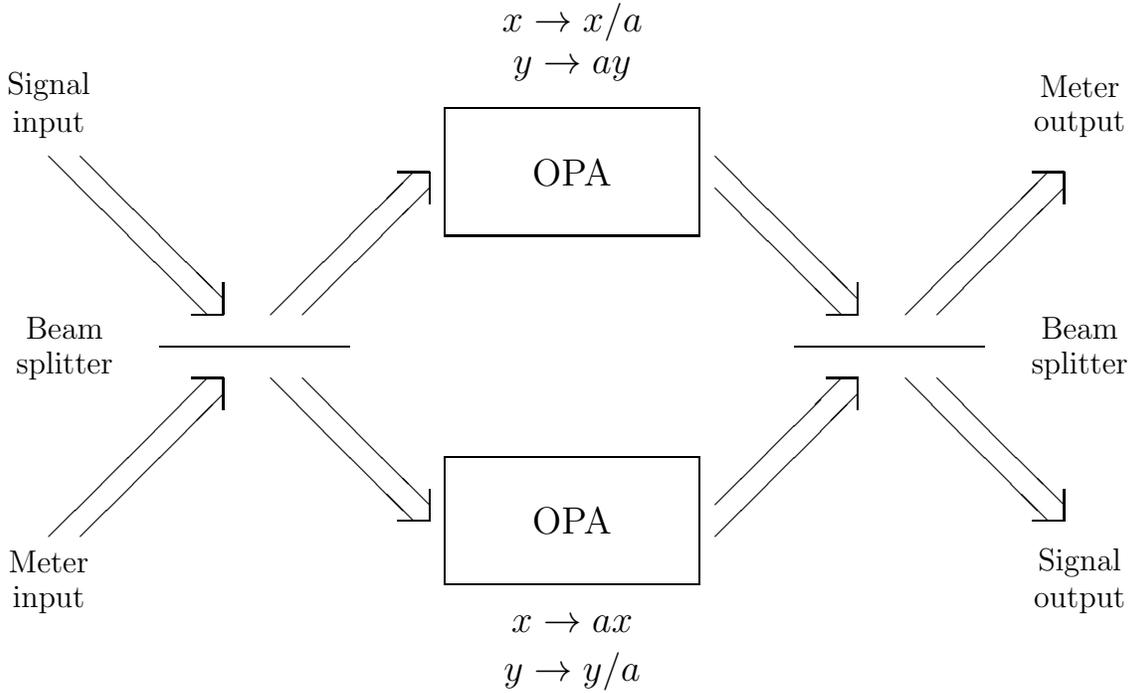
In this presentation, we investigate the creation of photons in the vacuum
by the backaction of a backaction evasion measurement of a quadrature 
component of the light field. An experimental setup for such a phase 
sensitive backaction evasion measurement was originally proposed 
by Yurke \cite{Yur85}. A schematic representation is shown in figure 
\ref{setup}. This measurement setup allows a measurement of the quadrature 
component $\hat{x}$ of the signal input. The precision of the measurement 
is limited by the vacuum fluctuations of the input meter field and the 
signal-meter coupling in the OPAs. For an amplification factor of $a$, 
the measurement resolution is given by $\delta\! x = a/2(a^2-1)$. 
The measurement statistics and the physical properties of the signal
output in this setup can be described by a generalized measurement operator,
\begin{equation}
\hat{P}_{\delta\!x}(x_m) = \left(2 \pi \delta\!x^2\right)^{-1/4} 
\exp \left(-\frac{(x_m-\hat{x})^2}{4\delta\!x^2}\right).
\end{equation}
Using this operator, the probability distribution $P(x_m)$ of 
measurement results $x_m$ and the quantum state
$\mid \psi_{\mbox{out}}(x_m)\rangle$ of the signal output for an 
arbitrary signal input state $\mid \psi_{\mbox{in}}\rangle$ can be determined 
by
\begin{eqnarray}
P(x_m) &=& \langle\psi_{\mbox{in}}\mid\hat{P}^2_{\delta\!x}(x_m)
            \mid  \psi_{\mbox{in}}\rangle
\nonumber \\
\mid \psi_{\mbox{out}}(x_m)\rangle &=& 
\frac{1}{\sqrt{P(x_m)}}\hat{P}_{\delta\!x}(x_m) \mid \psi_{\mbox{in}}\rangle.
\end{eqnarray}
The conditional final state $\mid \psi_{\mbox{out}}(x_m)\rangle$ defines
the probabilities for any further measurements performed on the signal 
field output. It is thus possible to determine joint probabilities
and correlations between the measurement result $x_m$ and 
other physical properties of the signal field.

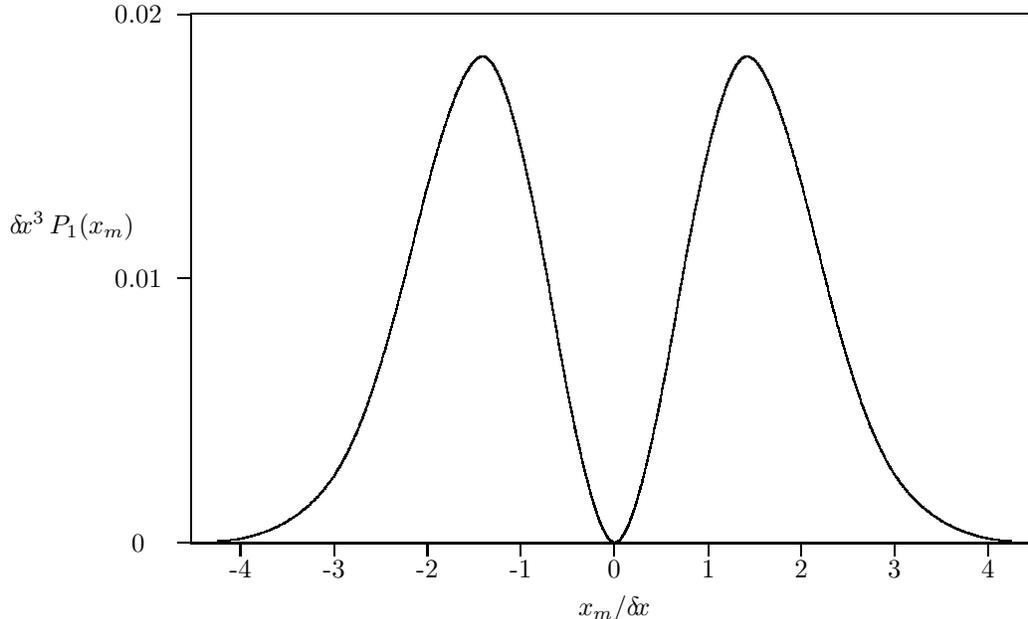
\begin{figure}[b]
\begin{picture}(470,240)
\put(80,30){\framebox(320,200){}}
\put(75,30){\line(1,0){5}}
\put(55,25){\makebox(10,10){0}}
\put(75,130){\line(1,0){5}}
\put(55,125){\makebox(10,10){0.01}}
\put(75,230){\line(1,0){5}}
\put(55,225){\makebox(10,10){0.02}}
\put(25,130){\makebox(20,40){$\delta\! x^3 \, P_1(x_m)$}}
\put(240,25){\line(0,1){5}}
\put(235,15){\makebox(10,10){0}}
\put(275.56,25){\line(0,1){5}}
\put(270.56,15){\makebox(10,10){1}}
\put(204.64,25){\line(0,1){5}}
\put(199.64,15){\makebox(10,10){-1}}
\put(310.7,25){\line(0,1){5}}
\put(305.7,15){\makebox(10,10){2}}
\put(169.3,25){\line(0,1){5}}
\put(164.3,15){\makebox(10,10){-2}}
\put(346,25){\line(0,1){5}}
\put(341,15){\makebox(10,10){3}}
\put(134,25){\line(0,1){5}}
\put(129,15){\makebox(10,10){-3}}
\put(381.4,25){\line(0,1){5}}
\put(376.4,15){\makebox(10,10){4}}
\put(98.6,25){\line(0,1){5}}
\put(93.6,15){\makebox(10,10){-4}}
\put(230,0){\makebox(20,10){$x_m/\delta\! x$}}
\bezier{300}(240,30)(248.4,30)(263.4,118)
\bezier{300}(263.4,118)(279.75,214)(290,214)
\bezier{200}(290,214)(299.3,214)(315.5,146.5)
\bezier{200}(315.5,146.5)(329.65,87.5)(340,66.5)
\bezier{200}(340,66.5)(355.4,32.5)(390,30.5)
\bezier{300}(240,30)(231.6,30)(216.6,118)
\bezier{300}(216.6,118)(200.25,214)(190,214)
\bezier{200}(190,214)(180.7,214)(164.5,146.5)
\bezier{200}(164.5,146.5)(150.35,87.5)(140,66.5)
\bezier{200}(140,66.5)(124.6,32.5)(90,30.5)
\end{picture}
\caption{\label{p1} Conditional probability $P_1(x_m)$ of finding a photon 
in the signal output following a measurement of $x_m$ for $\delta\! x \gg 1$.
The axes scale with $\delta\! x$.}
\vspace{0.5cm}
\end{figure}

In the case of a photon vacuum input state $\mid 0 \rangle$, the initial 
photon number is zero, but even low resolution ($\delta\!x>1$) measurements 
induce some quantum jumps to higher photon numbers. The discreteness of the 
quantum jump events make it difficult to understand the physical process by 
which the continuous field noise creates the photon. It is therefore 
interesting to examine the correlation between the continuous field 
measurement results $x_m$ and the quantum jump events. 
For $\delta\!x>1$, only zero and one photon contributions
are relevant, so we can focus on the joint probability of finding 
the one photon state $\mid 1 \rangle$ in the output signal after a 
measurement result of $x_m$ in the meter. In the limit of high $\delta\! x$,
this probability is given by
\begin{eqnarray}
P_1 (x_m) &=& 
|\langle 1 \mid \hat{P}_{\delta\!x}(x_m) \mid 0 \rangle |^2
\nonumber \\ &\approx&
\frac{1}{\sqrt{2\pi\delta\!x^2}} \frac{x_m^2}{(4\delta\!x^2)^2}
\exp\left(-\frac{x_m^2}{2\delta\!x^2}\right).
\end{eqnarray}
As shown in figure \ref{p1}, this is a double peaked probability 
distribution with peaks around $x_m=\pm \sqrt{2} \delta\!x$. 
The total area is equal to the quantum jump probability of 
$1/(16 \delta\!x^2)$. As $\delta\!x$ increases, the quantum jump 
probability decreases and the quantum jumps are associated with 
higher and higher field measurement results $x_m$.
This behavior is characterized by a constant correlation between 
quantum jump events and the squared measurement result $x_m$ given by
\begin{equation}
C(x_m^2;n)= \int d\!x_m P_1(x_m) \left( x_m^2 - \delta\!x^2 \right) 
= \frac{1}{8}.
\end{equation}
This constant correlation does not depend on the measurement resolution
and can therefore be interpreted as a fundamental physical property
of the vacuum state. In particular, an analysis of the measurement
operator $\hat{P}_{\delta\!x}(x_m)$ reveals that this correlation 
originates directly from the operator ordering dependence of products
of quantum variables. In terms of operator expectation values of the 
initial state, the quantum correlation observed in the measurement
described above reads 
\begin{equation}
C(\hat{x}^2,\hat{n}) = \frac{1}{4} \langle \hat{x}^2 \hat{n} + 
                  2 \hat{x}\hat{n}\hat{x} + \hat{n}\hat{x}^2 \rangle
          - \langle \hat{x}^2 \rangle \langle \hat{n} \rangle.
\end{equation}
Even though the vacuum state is an eigenstate of $\hat{n}$ with an
eigenvalue of zero, the operator correlation $C(\hat{x}^2,\hat{n})$
is nonzero because in the term $\langle \hat{x}\hat{n}\hat{x}\rangle$
the photon number operator is sandwiched between the quadrature component
operators. The (mathematical) action of $\hat{x}$ on $\mid 0 \rangle$
creates a one photon state $0.5 \mid 1 \rangle$. This creation of a 
one photon contribution is responsible for an operator correlation
of $C(\hat{x}^2,\hat{n})=1/8$ in the vacuum state.

The presence of a correlation between quadrature components
and photon number in an eigenstate of photon number raises fundamental
questions about the physical meaning of eigenvalues. Obviously, an 
eigenvalue is not just an ``element of reality'' as suggested in many
interpretations of quantum mechanics. The experimental
observation of correlations between the quantum jumps and the 
quadrature component measurement results could thus provide a new
perspective on the nature of quantum mechanical reality.
\\[0.5cm]
One of us (HFH) would like to acknowledge support from the Japanese
Society for the Promotion of Science, JSPS.


\begin{thebibliography}{}
\bibitem{Fri92} S.R. Friberg, S. Machida, and Y. Yamamoto, 
``Quantum-Nondemolition Measurement of the Photon Number of an 
Optical Soliton,'' Phys. Rev. Lett. {\bf 69}, 3165 (1992).

\bibitem{Per94} S.F. Pereira, Z.Y. Ou, and H. J. Kimble, ``Backaction 
Evading Measurements for Quantum Nondemolition Detection and Quantum 
Optical Tapping,'' Phys. Rev. Lett. {\bf 72}, 214(1994).

\bibitem{Hof00a}  H.F. Hofmann,
  ``Nonclassical correlations of phase noise and photon number in quantum
nondemolition measurements,'' Phys. Rev. A {\bf 61}, 033815 (2000).

\bibitem{Hof00b}  H.F. Hofmann, T. Kobayashi, and A. Furusawa,
  ``Nonclassical correlations of photon number and field components 
in the vacuum state,'' Phys. Rev. A  (to be published).

\bibitem{Yur85} B. Yurke, ``Optical back-action-evading amplifiers,''
J. Opt. Soc. Am. B {\bf 2}, 732 (1985).

\end{thebibliography}
\end{document}